\documentclass[english,aps,twocolumn,prb,showpacs,superscriptaddress]{revtex4}

\newcommand{\vecS}{\vec{\cal S}}
\newcommand{\sigb}{{\vec\sigma}}
\newcommand{\vecr}{\mathitbf r}
\newcommand{\ex}{\hat e_x}
\newcommand{\ey}{\hat e_y}

\DeclareMathAlphabet{\mathitbf}{T1}{cmr}{bx}{it}

\usepackage{epsfig,amsfonts,amssymb,newlfont,rotate,float,array}

\usepackage{t1enc,babel}
\usepackage{graphicx,bm}

\begin{document}

\title{Phase diagram and influence of defects in the
double perovskites}

\author{J.~L.~Alonso}
\affiliation{Departamento de F{\'\i}sica Te\'orica, Facultad de Ciencias,
Universidad de Zaragoza, 50009 Zaragoza, Spain.}
\affiliation{Instituto de Biocomputaci\'on y F\'{\i}sica de 
Sistemas Complejos (BIFI). Universidad de Zaragoza, 50009 Zaragoza, 
Spain.}
\author{L.~A.~Fern\'andez}
\affiliation{Departamento de F{\'\i}sica Te\'orica, 
Facultad de Ciencias F{\'\i}sicas,
Universidad Complutense de Madrid, 28040 Madrid, Spain.}
\affiliation{Instituto de Biocomputaci\'on y F\'{\i}sica de 
Sistemas Complejos (BIFI). Universidad de Zaragoza, 50009 Zaragoza, 
Spain.}
\author{F.~Guinea}
\affiliation{Instituto de Ciencia de Materiales (CSIC). Cantoblanco,
28049 Madrid. Spain.}
\affiliation{Instituto de Biocomputaci\'on y F\'{\i}sica de 
Sistemas Complejos (BIFI). Universidad de Zaragoza, 50009 Zaragoza, 
Spain.}
\author{F.~Lesmes}
\affiliation{Centro de Astrobiolog{\'\i}a. INTA-CSIC.
Carretera de Ajalvir km 4, 28850 Torrej\'on de Ardoz,
Madrid, Spain.}
\author{V.~Mart\'{\i}n-Mayor}
\affiliation{Departamento de F{\'\i}sica Te\'orica, 
Facultad de Ciencias F{\'\i}sicas,
Universidad Complutense de Madrid, 28040 Madrid, Spain.}
\affiliation{Dipartimento di Fisica,
Universit\`a di Roma ``La Sapienza'', INFN, SMC and UdR1 of INFM,
P.le Aldo Moro 2, 00185 Roma, Italy.}
\affiliation{Instituto de Biocomputaci\'on y F\'{\i}sica de 
Sistemas Complejos (BIFI). Universidad de Zaragoza, 50009 Zaragoza, 
Spain.}
\date{14th October 2002}
\begin{abstract}
The phase diagram of the double perovskites of the type 
Sr$_{2-x}$La$_x$FeMoO$_6$ is analyzed, with and without disorder due
to antisites. In addition to an homogeneous 
half metallic ferrimagnetic phase in the absence of doping and
disorder, we find antiferromagnetic phases at large dopings,
and other ferrimagnetic phases with lower saturation magnetization, in the
presence of disorder. 
\end{abstract}
\maketitle
{\it Introduction.}
The double perovskite Sr$_2$FeMoO$_6$ and related materials\cite{SW72}
are good candidates for magnetic devices, as they combine
a high Curie temperature and a fully polarized (half metallic)
conduction band.\cite{Ketal99} At present, these materials are
being extensively studied.\cite{Oetal99,Metal00,
Setal00,Getal00,Retal00,Netal01,
Betal01,Getal01,Metal01,Getal01b,Setal02,Setal02b,Tetal02}

The magnetism of these compounds arises from the Fe$^{3+}$, $S = 5/2$
core spin, while the charge state of the Mo ion is $5+$. Spatially,
the Mo and Fe ions occupy two interleaving FCC lattices (sodium
chloride structure).  The conduction band contains one electron per
unit cell, which tends to be antiparallel to the Fe spin. Experiments
suggest that, in many samples, the saturation magnetization is less
than the expected $4 \mu_B$ per formula unit.  This effect is usually
ascribed to the presence of antisite
defects,\cite{Oetal99,Setal00,Betal01,Metal01,
Netal01,Getal01,Setal02,Setal02b,Vetal02,SS01} where, due to the similarity of
their atomic radii, Mo ions are randomly placed on the Fe sublattice
and conversely. Notice that when a Fe ion is misplaced, with high
probability it will have a Fe ion among its first neighbors, enhancing
direct antiferromagnetic (AFM) superexchange with respect to the ideal
structure. The strength of this coupling can be inferred
from the compound LaFeO$_3$, which has the same structure, but 
where the Mo ions have been substituted by
Fe$^{3+}$.  LaFeO$_3$ is known to be AFM,\cite{RCA93}
with a N\'eel temperature of $T_\mathrm{N}= 720$K.

The Sr ions in Sr$_2$FeMoO$_6$ can be substituted for trivalent
cations, like La, leading to
Sr$_{2-x}$La$_x$FeMoO$_6$.\cite{Metal00,Netal01,Setal02} These
compounds have $1+x$ electrons per formula unit in the conduction
band.  These doped materials tend to have a higher Curie temperature.
Notice that one can also consider the substitution with a monovalent
ion ({\em i.e.} Sr$\rightarrow$ K, Sr$_{2-x}$K$_x$FeMoO$_6$ ), which
takes one electron from the conduction band, leaving $1-x$ electrons
per formula unit. Hence in this paper negative $x$ will actually refer
to substitution with a monovalent ion.

{\it The model.}
Band structure calculations have shown that the conduction band 
can be described in terms of hybridized ${\mathitbf t}_{2g}$ orbitals
at the Mo and Fe sites.\cite{Ketal99,Setal00b}
If one considers the ${\mathitbf t}_{2g}$ orbitals of both
spin orientations at the Fe sites, the model
leads to a highly correlated system, where an 
on site Hund's coupling and a Hubbard repulsive term
have to be added.\cite{CM01,Aetal01,PA02} In the following, we
will consider the magnetic phase diagram only, and neglect
the possible existence of a metal-insulator transition
when the ratio between the bandwidth and the Coulomb term
is sufficiently small.\cite{Aetal01,PA02} 
We consider that the conduction band is built up of the three 
${\mathitbf t}_{2g}$ orbitals at the Fe sites with spins oriented
antiparallely to the Fe moment, and the six
${\mathitbf t}_{2g}$ orbitals at the Mo sites (see below). 

We denote the destruction operator on $xy$ orbitals with spin $+$ or
$-$ at lattice site $\vecr$ as $F_{xy;\pm;\vecr}$, $M_{xy;\pm;\vecr}$
($F$ for Fe and $M$ for Mo), and so on. The spin and number operators
on a given Fe site are:
\begin{widetext}
\begin{eqnarray}
\vecS_\vecr&=&\sum_{\alpha,\beta=\pm} \left(F^\dag_{xy;\alpha;\vecr}+
F^\dag_{xz;\alpha;\vecr}+F^\dag_{yz;\alpha;\vecr}\right) \sigb_{\alpha,\beta}
\left(F_{xy;\beta;\vecr}+ 
F_{xz;\beta;\vecr}+F_{yz;\beta;\vecr}\right)\,,\\
{\cal N}^{\mathrm{Fe}}_\vecr&=&\sum_{\alpha=\pm}
(F^\dag_{xy;\alpha;\vecr}F_{xy;\alpha;\vecr}+
F^\dag_{xz;\alpha;\vecr}F_{xz;\alpha;\vecr}+
F^\dag_{yz;\alpha;\vecr}F_{yz;\alpha;\vecr})\,.
\end{eqnarray}
\end{widetext}
Analogous definitions hold for the Mo atoms. Given the large spin
value (S=5/2) of the the localized Fe core spins, we treat them as
classical, with polar coordinates
\begin{equation}
\vec\phi=(\sin \theta\, \cos\varphi,\sin\theta\, \sin\varphi, \cos\theta)
\end{equation}
As mentioned above, we only consider the Fe orbitals with spin
antiparallel to $\vec\phi$, which amounts to assume that the Hund's
coupling at the Fe ions is much larger than the other
interactions. Thus, we define up and down orbitals, $f_1$ and $f_2$,
with respect to the local 5/2 spin:
\begin{eqnarray}
F_+&=& \cos \frac{\theta}{2} f_1 + \sin \frac{\theta}{2} f_2\,,\nonumber \\
F_-&=& \sin \frac{\theta}{2}e^{\mathrm i\varphi} f_1 - \cos
\frac{\theta}{2} e^{\mathrm i\varphi} f_2\,,
\end{eqnarray}
and we neglect all terms including the $f_1$ operators. For the sake
of brevity, in the following we set $f_2 = f$. Then, the Hamiltonian,
in the absence of disorder and neglecting direct hopping terms between
Mo orbitals (see below), can be written as:
\begin{equation}
{\cal H}= {\cal K}_{xy}+ {\cal K}_{yz} + {\cal K}_{xz}
 -\mu\!\sum_{\vecr\,\mathrm{even}\atop} {\cal N}^{\mathrm{Fe}}_\vecr-
 (\mu+\Delta)\!\!\sum_{\vecr\, \mathrm{odd}\atop} {\cal N}^{\mathrm{Mo}}_\vecr\!,
\label{hamil}
\end{equation}
with
\begin{equation}
{\cal K}_{xy} 
= t_\mathrm{Mo-Fe}\!\!\!\sum_{{\vecr\in \mathrm{Fe\, lattice}}\atop{\hat u=\ex,\ey}}
\!\!(\sin\frac{\theta_\vecr}{2}f^\dag_{xy;\vecr}
M_{xy;+;\vecr+\hat u}+\mathrm{h.c.})\, ...
\end{equation}
where, for brevity, we omit the analogous hoppings from the sites
belonging to the Mo sublattice to the Fe sites.  Analogous expressions
are found for the kinetic energy on the $xz$ and $yz$ planes. Finally,
we add a direct hopping between Mo orbitals.  These hopping terms give
rise to three separate two dimensional
Hamiltonians.\cite{CM01,Aetal01,PA02} The substitution of Mo ions for
Fe ions leads to direct Fe-Fe hopping terms, ( we take
$t_\mathrm{Fe-Fe} = t_\mathrm{Mo-Fe}$), and also to the inclusion of
an AFM exchange, $J_\mathrm{Fe-Fe}$. Thus, the model is
defined by the parameters $t_\mathrm{Mo-Fe} , t_\mathrm{Mo-Mo} ,
\Delta , \mu$ and $J_\mathrm{Fe-Fe}$.  There are nine orbitals per
unit cell, three at the Fe sites, and six at the Mo sites.

The occupancy of the conduction band depends on the value of
the chemical potential, $\mu$, and it
varies from one electron to two electrons per unit cell in
Sr$_{2-x}$La$_x$FeMoO$_6$, $0 \le x \le 1$. We neglect
interactions of the electrons within this band (as discussed
below, the number of electrons at the Fe sites is always less
than one). 

We will use $t_\mathrm{Fe-Mo}$ as our unit of energy
($t_\mathrm{Fe-Mo} \approx 0.35$eV from band structure
calculations). We take $t_\mathrm{Mo-Mo} / t_\mathrm{Fe-Mo} = 0.25$,
$\Delta = 0$ and $J_\mathrm{Fe-Fe} / t_\mathrm{Fe-Mo} = 0.1$. The
value of $\Delta $ implies a relatively large hybridization of the Fe
and Mo orbitals, which seems consistent with Hartree-Fock
calculations.\cite{PA02} $J_\mathrm{Fe-Fe}$ is chosen so as to
reproduce the N\'eel temperature of LaFeO$_3$. We have not made a
comprehensive study of the dependence of the results on the tight
binding parameters, but the calculations made so far indicate that the
qualitative features of the phase diagrams to be discussed below are
not strongly dependent on the choice of parameters.  In the absence of
disorder, this model is basically equivalent to the one studied by
Chattopadhyay and Millis~\cite{CM01} in the context of Dynamical Mean
Field Theory, although we shall use Variational Mean Field (see
\onlinecite{Aetal01a} for a comparison between the two methods). The
main novelty is in our considering of the disorder
effects\cite{Oetal99,Setal00,Betal01,Metal01,
Netal01,Getal01,Setal02,Setal02b,SS01}: with probability $y$ we
misplace an Fe ion onto the Mo sublattice (and conversely) without any
spatial correlations ($y$ is just the antisite density). It is clear
that $y=0.5$ corresponds to full disorder on the location of the Fe
and Mo ions, while $y>0.5$ is equivalent to $1-y$ with the Fe and Mo
sublattices interchanged. Vacancies can be equally considered, but
explicit calculations showed that they have a much milder effect on
the phase diagram.

{\it Method of calculation.}
We use the method developed for double exchange systems 
in Ref. \onlinecite{Aetal01a}. We assume that the Fe core spins are classical. 
At a given temperature, we average over spin configurations
obtained by assuming that there is a magnetic field acting on
the spins. The magnitude of these fields are variational parameters,
which are taken so as to minimize the free energy. 
Given a spin configuration, the electronic states are calculated
exactly, and the electronic contribution to the free energy is obtained
by integrating the density of states. As the Fe spins are distributed
in a three dimensional lattice, and the electrons lead to effective
interactions with the cubic symmetry, we think that our mean field ansatz
for the spin configurations is sufficient. 
This method is in excellent agreement with more
precise Monte Carlo calculations for the double
exchange model.\cite{Aetal01b} We solve the Hamiltonian in lattices
with up to $512 \times 512 \times 512$ sites (note that
the calculation of the electronic wave functions requires
only the diagonalization of the Hamiltonian in a
$512 \times 512$ square).  For these sizes,
the disorder due to antisites is self-averaging.

The adequacy of our technique depends on the ans\"atze made for the
possible spin configurations. We have considered four possible
phases: i) the paramagnetic (PM) phase, ii) the ferrimagnetic (FI) phase, where
all Fe spins are parallel, and the spins of the electrons in the conduction 
band are antiparallel to the Fe spins,
iii) an AFM phase, where
the Fe spins in neighboring (1,1,1) planes are antiparallel,
and iv) a different 
ferrimagnetic (FIP) phase where the Fe spins are aligned 
ferromagnetically if the Fe are in the correct positions, and
antiferromagnetically if the Fe ions occupy Mo sites because
of the antisite defects. In the absence of disorder, we have checked
that other phases with canted spins have higher
free energy. Note that the above ans\"atze define the average
magnetization at the Fe sites, but that thermal fluctuations are also
included.

\begin{figure}
\includegraphics[angle=90,width=\columnwidth]{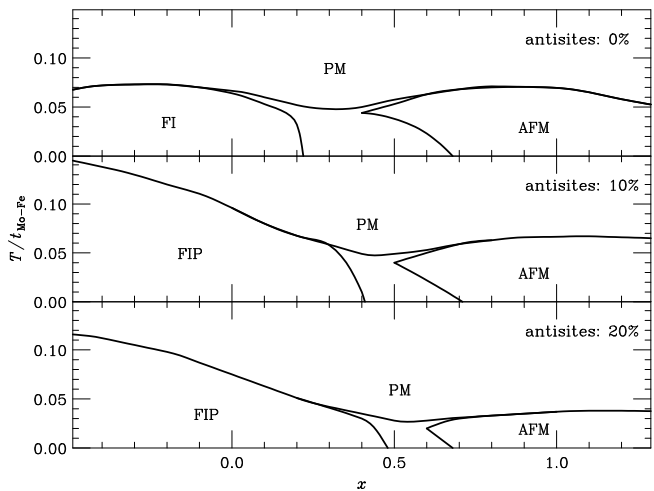}
\caption{Phase diagram of Sr$_{2-x}$La$_x$FeMoO$_6$ as function of $x$
and temperature for different concentrations of antisite
defects. Negative $x$ actually means
Sr$_{2-|x|}$K$_{|x|}$FeMoO$_6$. In both cases, the density of
carriers in the conduction band is $1+x$.  Phase-separation regions
are found between the FI and AFM phases (upper panel), and between the
FIP and AFM phases (middle and lower panel).}
\label{phase_d}
\end{figure}

{\it Results.}
The phase diagram of Sr$_{2-x}$La$_x$FeMoO$_6$, as function of
$x$ and temperature, is shown in Fig.~\ref{phase_d} for different
concentrations of antisites.

In the absence of defects, we find that $T_{\mathrm C}$ decreases
with increasing doping of the conduction band, in agreement
with Ref.~\onlinecite{CM01}. 
At high, but still reasonable, dopings we find the ordered
AFM phase described above. The phase transitions
are first order, with regions of phase separation between them.
For $x \approx 0$, the spins of the electrons at the Mo orbitals
are antiparallel to the Fe core spins.
We ascribe the tendency toward phases with zero
magnetization, upon increasing doping,
to the occupancy of the Mo orbitals which are
aligned parallel to the Fe spins.

The presence of antisite defects changes significantly the phase
diagram: i) The FI phase is replaced by the FIP phase, where the spins
at the Fe sites at the defects are antiparallel to the overall
magnetization, ii) the ordered AFM phase is strongly suppressed, and
iii) the value of $T_{\mathrm C}$ increases as the concentration of
antisites also increases~\cite{estimate}, iv) the dependence of
$T_{\mathrm C}$ with the number of electrons in the conduction band is
more pronounced in the presence of antisites.

These effects are associated to the direct AFM
interaction between spins at Fe ions which are nearest neighbors.
These interactions play no role in perfect materials. The
antiferromagntic interaction can be easily shown to be equivalent to a
{\em ferromagnetic} one for the atoms in the Fe sublattice.  Thus,
superexchange enhances the tendency toward a ferromagnetic order in
the original Fe sublattice.  This effect is independent of the number
of electrons in the conduction band. The saturation magnetization, on
the other hand, is reduced.

\begin{figure}
\includegraphics[angle=90,width=\columnwidth]{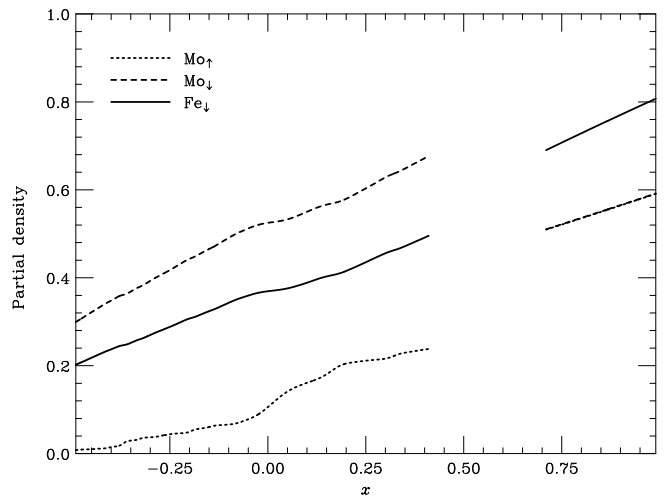}
\caption{Occupation of the Mo$_\uparrow$, Mo$_\downarrow$ and
Fe$_\downarrow$ as function of the doping of the conduction band.  The
curves give the occupancies for a $10\%$ density of antisites defects.
Note that in phases with no net magnetization, the occupancies of the
Mo$_\uparrow$ and Mo$_\downarrow$ levels are the same.}
\label{dos_int}
\end{figure}
Fig.~\ref{dos_int} gives the occupancies of the different
orbitals as the number of electrons in the conduction band
is varied. Most of the charge is in the Mo orbitals.
The variation is not linear, indicating that a rigid band
picture is not valid.\cite{SS01} There are sharp changes
at the phase transitions.

\begin{figure}
\includegraphics[angle=90,width=\columnwidth]{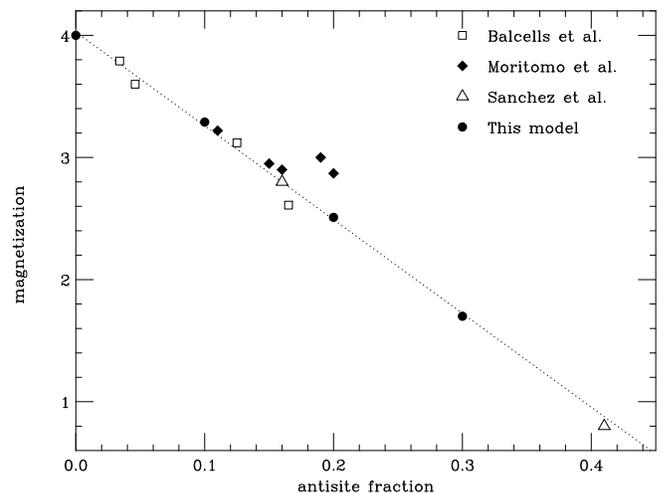}
\caption{Low temperature magnetization ($\mu_\mathrm{B}$ per formula unit) 
of Sr$_2$FeMoO$_6$
as function of the concentration of antisite defects, $y$.
Experimental results are from Refs.~\protect{\onlinecite{Betal01,Metal01,
Setal02b}}.}
\label{magnetization}
\end{figure}

At low temperatures, the spins at antisites tend to be
antiparallel to the magnetization, as shown 
in Fig.~\ref{phase_d}. 
This implies that the saturation
magnetization is reduced with respect to the ordered
case. The total magnetization 
of the core spins and the conduction electrons,
is shown in Fig.~\ref{magnetization}. 
The calculated magnetization is well fitted by 
the line $M_S = (4.0 - 7.7 y) \mu_B$, where  $y$ is the antisite density.
Experimental results from Refs.~\onlinecite{Betal01,Metal01,
Setal02b}. are added for comparison. Note that the decrease in the
magnetization  does not lead to a lowering of the Curie temperature, 
as discussed above.

{\it Conclusions.}
We have studied the magnetic phase diagram of the doped double perovskites,
Sr$_{2-x}$La$_x$FeMoO$_6$. We have analyzed the influence of antisite defects,
on the phase diagram.

In clean systems, we find that, as the number of electrons in the
conduction band increases, the critical temperature decreases, 
in agreement with previous calculations.\cite{CM01} This variation 
is due to the increased filling of the Mo$_\uparrow$ band, which 
reduces the double exchange-like  mechanism which tends to align the Fe
moments. At sufficiently high dopings, we find ordered phases
without net magnetization, which  enhance the delocalization
of both the Mo$_\uparrow$ and Mo$_\downarrow$ bands. 
The transitions
between these phases tend to be first order, with regions
of phase separation between them. Electrostatic effects will
prevent the existence of phase separation at macroscopic scales,
leading to a domain structure at mesoscopic scales\cite{AGG00}.

Antisite disorder induces significant changes in the phase diagram.
The ordered ferrimagnetic phase is replaced by
a different ferrimagnetic phase where the Fe spins at defects
are antiparallel to the bulk magnetization
(the FIP phase, see Fig.\ref{phase_d}). Antiferromagnetism at
finite dopings is suppressed.
The saturation magnetization in the FIP phase is reduced,
although
the Curie temperature tends to increase
with the number of Fe in Mo positions, due to the direct
AFM exchange between Fe ions which are
nearest neighbors\cite{estimate}.

Note that,
in order to study compounds with different number
of carriers in the conduction band, the presence
of vacancies and changes in the Fe - O - Mo
bond angles, Fe/Mo - O distance, and in the energy splitting 
$\Delta$ can influence the results. These effects need to
be extracted from the available experimental data and incorporated
in the model Hamiltonian, eq.~(\ref{hamil}). 

We have not studied transport properties, although it seems likely
that the variation of the magnetic structure near defects
will lead to significant changes in a half metallic system.\cite{SS01}
We have also not analyzed other effects of the
electron-electron interaction, such as the 
existence of a Mott transition to an insulating state,
found in the related compound Sr$_2$FeWO$_6$.\cite{PA02}
We think, however, that our model includes all relevant
interactions required to study the magnetic properties of
the metallic state of double perovskites. Similar models
provide a good understanding of the magnetic 
properties of the half metallic manganite oxides
(such as La$_{1-x}$Ca$_x$MnO$_3$).\cite{Aetal01a,Aetal01b}

In summary, we find a rich phase diagram for 
Sr$_{2-x}$La$_x$FeMoO$_6$, which is significantly modified
in the presence of defects. Our results seem consistent
with existing experimental data.

{\it Acknowledgements.}  We are thankful to J. Blasco, J.M.D. Coey,
J.M. de Teresa, J. Fontcuberta, M. Garc{\'\i}a-Hern\'andez,
M.R. Ibarra, J. L.  Mart{\'\i}nez, L. Morell\'on, D. D. Sarma, and
D. Serrate, M. Venkatesan, and especially to V. Laliena, for helpful
discussions. VMM was supported by E.C. contract HPMF-CT-2000-00450 and
by a {\em Ramon y Cajal} contract (Spain).  Financial support from
grants PB96-0875, FPA2000-0956, FPA2000-1252, FPA2001-1813 (MCyT,
Spain), and 07N/0045/98 (C. Madrid) are acknowledged.


\begin{thebibliography}{99}
\bibitem{estimate}
Our estimate of $J_\mathrm{ Fe-Fe}
/ t_\mathrm{ Fe-Mo} = 0.1$ gives a lower bound to
this anomalous dependence of T$_\mathrm{C}$
on the number of defects. A lower, but physically reasonable, 
value for $t_\mathrm{Fe-Mo}$ than the
one used here, $t_\mathrm{ Fe-Mo} \approx 0.35$eV, leads to
a more pronounced increase 
of T$_\mathrm{C}$ as function of
the number of antisite defects, for fixed
doping.
\end{thebibliography}
\end{document}